\documentclass[conference,twocolumn]{IEEEtran}

\usepackage{amsmath}
\usepackage{amsthm}
\usepackage{amssymb}
\usepackage{mathtools}
\usepackage{graphicx}
\usepackage{rotating}
\usepackage{enumitem}
\usepackage{tcolorbox}
\usepackage{xcolor}
\usepackage{tikz}

\newsavebox{\tempbox}

\DeclareMathAlphabet\mbcf{OMS}{cmsy}{b}{n}

\newcommand{\logt}{\log_{2}}

\newcommand{\defn}{\triangleq}

\newcommand{\mcf}[1]{\mathcal{#1}}

\newcommand{\set}[1]{\left\{ #1 \right\}}

\newtheorem{theorem}{Theorem}
\newtheorem{define}[theorem]{Definition}

\newtheorem{lemma}[theorem]{Lemma}

\newtheorem{remark}[theorem]{Remark}
\newtheorem{convention}[theorem]{Convention}

\newenvironment{proofsketch}{%
  \proof}{\endproof}

\let\oldremark\remark
\renewcommand{\remark}{\oldremark\normalfont}

\IEEEoverridecommandlockouts

\usepackage{etoolbox}

\providetoggle{arxiv}
\settoggle{arxiv}{true}

\begin{document}

\title{Modular design to transform codes for the wiretap channel of type I into codes for the wiretap channel of type II}

\author{\IEEEauthorblockN{Eric Graves}
\IEEEauthorblockA{Army Research Lab\\
Computer and Information Sciences Division \\
Adelphi, MD. 20783\\
Email: ericsgraves@gmail.com}
\and
\IEEEauthorblockN{Allison Beemer}
\IEEEauthorblockA{Arizona State University\\
School of Electrical, Computer and Energy Engineering\\
Tempe, AZ 85287\\
Email: allison.beemer@asu.edu}}
\maketitle

\begin{abstract}
We construct a modular scheme which extends codes for wiretap channels of type I for use in wiretap channels of type II.
This is done by using a concatenate and permute strategy, wherein multiple uses of the wiretap type I code are concatenated and then the joint sequence of symbols permuted. 
The choice of permutation is then encoded with a short code and appended to the transmitted sequence.
Analysis shows essentially no degradation in operational parameters (rate, error rate, leakage) for the new code over the wiretap type II channel when compared to those of multiple uses of the original code over the wiretap type I channel.

\end{abstract}

\section{Introduction}

Wyner~\cite{wyner75wtc}, and later Csisz{\'a}r and K{\"o}rner~\cite{csiszar1978broadcast}, first studied the wiretap channel of type I, which is a model of the communication scenario where an encoder wants to send private information to a decoder in the presence of an eavesdropper. 
Specifically, Wyner considered the discrete memoryless case, where the channel from encoder to eavesdropper was a degraded version of the channel from encoder to decoder, while Csisz{\'a}r and K{\"o}rner considered a slightly more general model which included the case where the channel from encoder to eavesdropper no longer need be degraded.
In both cases, the statistical descriptions of both channels were fixed and known by all parties.
The result of the analysis determined the maximum bits per symbol at which an encoder could send the information, and still have the mutual information per bit between the message and the eavesdropper's observation be small.

Of course, this model does not accurately reflect real world scenarios; it is unlikely for the channel from encoder to eavesdropper to be known by the encoder, since learning a channel generally requires cooperation from both parties.
For this reason, alternative wiretap channel models were formulated, this time with some amount of ambiguity on the part of the encoder about the channel from encoder to eavesdropper. 
The wiretap channel of type II, introduced by Ozarow and Wyner~\cite{ozarow1984wtcII}, is one such model.
In the original wiretap channel of type II, the channel from encoder to decoder could be used to perfectly send any length-$n$, $n \in \mathbb{Z}^+$, sequence, but the eavesdropper could perfectly observe any $k, 0<k< n$, symbols of their choice. 
Here the ambiguity entered by way of the encoder not knowing which of the symbols the eavesdropper would choose. 

Recently, Nafea and Yener~\cite{nafea2015wiretap,nafea2018new} generalized the wiretap channel of type II by allowing the channel from the encoder to decoder to be noisy, and allowing multiple noisy channels from encoder to eavesdropper which, for every symbol, the eavesdropper may select to use up to a set number of times.
Once again, this model was thought to be closer to a realistic scenario since the encoder is not allowed to know the eavesdropper's choice of channel (instead only which channels are possible, and the number of times that channel could be chosen) a priori. 
Not long after the new model had been proposed, Goldfeld et al.~\cite{goldfeld2016semantic} determined the semantic secrecy capacity for the case where the channels from the encoder to eavesdropper are either perfect or convey no information. 
Their result stands primarily as a demonstration of the power of Cuff's~\cite{cuff2016soft} soft-covering lemma.
Similar results for the more general wiretap channel of type II were also obtained for strong secrecy by He et al.~\cite{he2016strong} and by Nafea and Yener~\cite{nafea2018new}.

The purpose of our work is to demonstrate that nearly any code for the wiretap channel of type I has a modular extension for wiretap channels of type II.
In other words, it is somewhat unnecessary to design a purpose built code for a wiretap channel of type II: codes for the wiretap channel type I will suffice with some minor modification. 
It also suggests that most results for the wiretap channel of type I directly extend to a wiretap channel of type II.
To do this we exploit the fact the eavesdropper must choose their state sequence independently of the message. 
This exploitation comes in the form of applying Ahlswede's~\cite{ahlswede1978elimination} ``robustification process'' to multiple uses of a wiretap channel type I code, while sending the randomization used for the robustification via a shorter code.
Doing so, the robustification process removes the ability for the eavesdropper to correlate their channel state selection with the code in a meaningful way.
The only degradation in the operational parameters comes in that the wiretap type I code had to be used multiple times. 
Comparing the new wiretap channel type II code to multiple uses of the wiretap channel type I code sees nearly equivalent measures of operation.

% Nafea and Yener~\cite{nafea2016new,nafea2016mac,nafea2016macII,nafea2017ibccc,nafea2017bc}. Other junk from them

% Tahamsbi et al.~\cite{tahamasbi2017learning}

% Maurer~ strong leakage requirement \cite{maurer1994strong}

% Goldfeld's work on arbitrarily varying wiretap channels \cite{goldfeld2016arbitrarily} 

% Bellare, et al. \cite{bellare2012cryptographic,bellare2012semantic} semantic leakage

\section{Model and notation}

\subsection{Notation}
Random variables, constants, and sets will be written with upper case, lower case, and script respectively.
For example $X$ may take on value $x\in \mcf{X}$. 
$X_j^n$ will be used to denote the sequence of random variables $X_j,X_{j+1},\dots,X_n.$ If $j =1,$ the subscript will be omitted. 
Similar notation will be used to denote sequences of constants and sets. 

$\mcf{P}(\mcf{X})$ will be used to denote the set of all probability distributions on a discrete set $\mcf{X}$, similarly $\mcf{P}(\mcf{Y}|\mcf{X})$ will be used to denote the set of conditional probability distributions on $\mcf{Y}$ given an element in $\mcf{X}$. 
Next $\mcf{P}_n(\mcf{X})$ denotes the set of all possible \emph{empirical distributions} over $\mcf{X}^n$.
Here an empirical distribution of a sequence is the normalized count of symbol occurrence. For instance, the sequence $1,0,0,1,a,1$ drawn from $\mcf{X} = \{0,1,a\}$ has empirical distribution $q(0) = \frac{1}{3},$ $q(1) = \frac{1}{2}$, $q(a) = \frac{1}{6}.$
Note, for any $p \in \mcf{P}_{n}(\mcf{X})$, $p(x)n \in \mathbb{Z}.$
Furthermore $\mcf{T}_{p}^n$ denotes the $n$-symbol type set of $p$, or in other words the set $\mcf{\hat X} \subset \mcf{X}^n$ containing all $x^n$ with empirical distribution $p$.
When necessary, we shall use $p_{X}$ to refer to the probability distribution over $X$, that is $p_{X}(x) = \Pr \left(X = x\right).$
Finally we shall use $p^n(x^n)$ to mean $\prod_{i=1}^n p(x_i).$

%For arbitrary products of probability distributions we shall use $\cdot,$ for instance given $p \in \mcf{P}(\mcf{X})$ and $ q \in \mcf{P}(\mcf{Y})$ then $p \cdot q$ is the distribution $p(x)$

To discuss the average of certain functions of random variables we shall use math blackboard bold font. 
In particular for random variables $X,Y,Z$ we will employ 
\begin{align*}
\mathbb{E}\left[ X \right] &= \sum_{x} p_{X}(x) x, \\
\mathbb{H}(X|Y)&\defn - \sum_{x,y} p_{X,Y}(x,y) \log p_{X|Y}(x|y),\\
\mathbb{I}(X;Y|Z)&\defn  \sum_{x,y,z} p_{X,Y,Z}(x,y,z) \log  \frac{p_{X,Y|Z}(x,y|z)}{p_{X|Z}(x|z) p_{Y|Z}(y|z)}.
\end{align*}
At one point in the paper it will be necessary to consider the last equation as a random variable depending on the value of $Z,$ thus we make note of this now
$$\mathbb{I}_{Z}(X;Y) = \sum_{z\in \mcf{Z}} 1\{ Z=z \} \mathbb{I}(X;Y|Z=z).$$

We will also speak of $n$-symbol sequence permutations.
For example, letting $\upsilon^*(\{1,2,3\}) = \{3,1,2\}$, then  $$\upsilon(x_1,x_2,x_3) = x_{\upsilon^*(1)},x_{\upsilon^*(2)},x_{\upsilon^*(3)} = x_3,x_1,x_2$$
is a $3$-symbol sequence permutation.

\subsection{Model}

\tikzstyle{block} = [draw, fill=white, rectangle, 
    minimum height=30pt, minimum width=30pt, text centered]

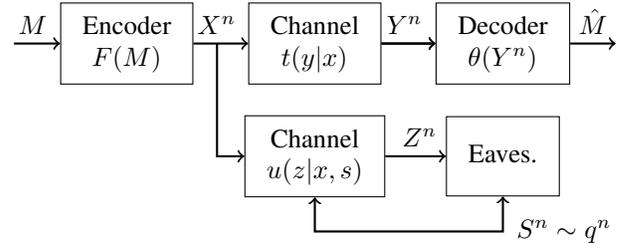
\begin{figure}
    \centering
\begin{tikzpicture}
\node[block] (enc) at (-.5,0) {$\begin{array}{c} \text{Encoder} \\ F(M) \end{array}$};
\node[block] (chan) at (2,0) {$\begin{array}{c} \text{Channel} \\ t(y|x)\end{array}$};
\node[block] (chanz) at (2,-1.5) {$\begin{array}{c} \text{Channel} \\  u(z|x,s)  \end{array}$};
\node[block] (dec) at (4.5,0){$\begin{array}{c} \text{Decoder} \\ \theta(Y^n)  \end{array}$};
\node[block] (eaves) at (4.5,-1.5){$\begin{array}{c} \text{Eaves.}  \end{array}$};
\draw[->,thick] (-2,0) -- node[above]{$M$} (-1.5,0)  -- (enc.west) ;
\draw[->,thick] (enc.east)  node[above]{$~~~~~X^n$}  -- (chan.west) ;
\draw[->,thick] (chan.east)  node[above]{$~~~~~Y^n$}  -- (dec.west) ;

\draw[->,thick] (.7,0) -- (.7,-1.5) -- (chanz.west) ;
\draw[->,thick] (chan.east) -- (dec.west) ;
\draw[->,thick] (dec.east) -- node[above]{$\hat M$} (6,0) ;
\draw[->,thick] (chanz.east) -- node[above]{$\hspace{10pt} Z^n$} (3.5,-1.5) --(eaves.west);
\draw[<->,thick] (eaves.south)-- (4.5,-2.5) -- (4,-2.5) node{$\hspace{75pt} S^n \sim q^n  $} -- (2,-2.5)-- (chanz.south);

\end{tikzpicture}
 \caption{Wiretap type I} 
\label{fig:I}
\end{figure}

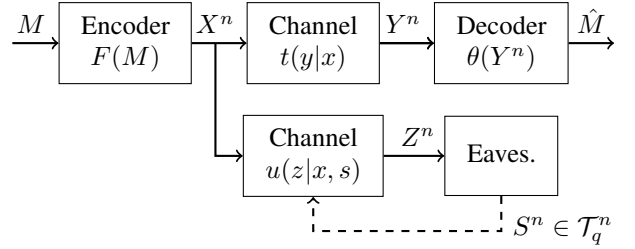
\begin{figure}
    \centering
\begin{tikzpicture}
\node[block] (enc) at (-.5,0) {$\begin{array}{c} \text{Encoder} \\ F(M) \end{array}$};
\node[block] (chan) at (2,0) {$\begin{array}{c} \text{Channel} \\ t(y|x)\end{array}$};
\node[block] (chanz) at (2,-1.5) {$\begin{array}{c} \text{Channel} \\ u(z|x,s) \end{array}$};
\node[block] (dec) at (4.5,0){$\begin{array}{c} \text{Decoder} \\ \theta(Y^n)  \end{array}$};
\node[block] (eaves) at (4.5,-1.5){$\begin{array}{c} \text{Eaves.}  \end{array}$};
\draw[->,thick] (-2,0) -- node[above]{$M$} (-1.5,0)  -- (enc.west) ;
\draw[->,thick] (enc.east)  node[above]{$~~~~~X^n$}  -- (chan.west) ;
\draw[->,thick] (chan.east)  node[above]{$~~~~~Y^n$}  -- (dec.west) ;
\draw[->,thick] (.7,0) -- (.7,-1.5) -- (chanz.west) ;
\draw[->,thick] (chan.east) -- (dec.west) ;
\draw[->,thick] (dec.east) -- node[above]{$\hat M$} (6,0) ;
\draw[->,thick] (chanz.east) -- node[above]{$\hspace{10pt} Z^n$} (3.5,-1.5) --(eaves.west);
\draw[->,thick,dashed] (eaves.south) -- (4.5,-2.5) -- (4,-2.5) node{$\hspace{75pt} S^n \in \mcf{T}_q^n $} -- (2,-2.5)-- (chanz.south) ;
\end{tikzpicture}
 \caption{Wiretap type II}
\label{fig:II}
\end{figure}

We begin by describing the wiretap type I channel model (depicted in figure~\ref{fig:I}), and then describing the differences between it and the generalized wiretap type II (depicted in figure~\ref{fig:II}).
In order to assist in presentation, the wiretap type I channel model here will be presented in a way to easier draw parallels between it and the wiretap channel of type II; it should be easy to see that this model is equivalent to the traditional channel model.

A wiretap type I channel model consists of an encoder, decoder and an eavesdropper. 
The encoder is a random function $F: \mcf{M} \rightarrow \mcf{X}^n$, which outputs a sequence $X^n = F(M)$ for message $M$.
This $n$-symbol sequence passes through a pair of communication channels to the decoder and eavesdropper who receive $Y^n$ and $Z^n$ respectively.
The communication channel between the encoder and decoder is described by a conditional probability distribution $t \in \mcf{P}(\mcf{Y}|\mcf{X})$ where given the encoder outputs $x^n$ the probability that the decoder receives $y^n$ is $\prod_{i=1}^n t(y_i|x_i)$.
On the other hand, the communication channel between the encoder and eavesdropper is a channel with state described by a conditional probability distribution $u \in \mcf{P}(\mcf{Z}|\mcf{X},\mcf{S})$, where given the encoder outputs $x^n$ and the state sequence is $s^n$ the probability that the eavesdropper receives $z^n$ is $\prod_{i=1}^n u(z_i|x_i,s_i)$.
The state sequence is generated i.i.d. according to $q\in \mcf{P}(\mcf{S})$, and is given the eavesdropper. 
The decoder $\theta : \mcf{Y}^n \rightarrow \mcf{M} $ then produces the message estimate $\theta(Y^n)$. 
We shall refer to a wiretap type I channel of the above form as WTC-I$(t,u,q)$.
\begin{define}
A pair $(F,\theta)$ is a $(n,r,\epsilon,\delta)$-code for the WTC-I$(t,u,q)$ if for $M$ uniformly distributed on $\mcf{M}$ the following are satisfied:
\begin{itemize}
    \item (Rate, Blocklength) $$|\mcf{M}| = 2^{nr}$$
    \item (Error probability) $$|\mcf{M}|^{-1} \sum_{m\in \mcf{M}} \Pr \left( \theta( Y^n) = m \middle| X^n = F(m) \right)  \geq 1- \epsilon,$$
    \item (Secrecy) $$\mathbb{I}(Z^n,S^n;M) = \mathbb{I}(Z^n;M|S^n) \leq \delta.$$ 
\end{itemize}
\end{define}

The wiretap type II channel is similar to the wiretap channel of type I, but now the wiretapper may choose the state $S^n$ subject $s^n \in \mcf{T}^n_{q}$ for a given $q \in \mcf{P}_n(\mcf{S})$. 
We shall refer to a wiretap type II channel of the above form as a WTC-II$(t,u,q)$.
\begin{define}
A pair $(F,\theta)$ is a $(n,r,\epsilon,\delta)$-code for the WTC-II $(t,u,q)$ if for $M$ uniformly distributed on $\mcf{M}$ the following are satisfied:
\begin{itemize}
   \item (Rate, Blocklength) $$|\mcf{M}|=2^{nr}$$, 
    \item (Error probability) $$|\mcf{M}|^{-1} \sum_{m\in \mcf{M}} \Pr \left( \theta( Y^n) = m \middle| X^n = F(m) \right)  \geq 1- \epsilon,$$
    \item (Secrecy) $$\max_{s^n \in \mcf{T}^n_{q}} \mathbb{I}(Z^n;M|S^n = s^n) \leq \delta.$$ 
\end{itemize}
\end{define}

\section{Results}

%\begin{lemma}\label{lem:type_to_iid}
%Let $n, \ell \in \mathbb{Z}^+$, and let $q \in \mcf{P}_n(\mcf{S})$.
%\[
%\frac{q^{n (\ell+1) } \left( \set{ s^{n (\ell+1)} : s^{n} = \hat s^{n}   } \cap \mcf{T}_{q}^{n(\ell+1)} \right)}{q^{n(\ell+1)}(\mcf{T}^{n(\ell+1)}_q)} \leq \mu q^{n}(\hat s^{n})
%\]
%where
%\begin{align*}
%\mu &= \left( 1 + \frac{2}{12 \phi -1} \right) \sqrt{\frac{\phi}{n} } \prod_{j\in \mcf{S}} a(j) \\
%a(j) &=  \max \left( e^{- \phi q(j)^2}  \sqrt{2 \pi n} , ~e^{-\frac{\zeta(j) }{8}} \sqrt{\frac{2}{1 + \sqrt{1 - 2\zeta(j) }}} \sqrt{\frac{n}{\phi}}   \right), \\
%\zeta(j) &= \frac{n_1^2}{\phi^3 q(j)^2}, \\
%\phi &= n\ell  .
%\end{align*}
%end{lemma}

Our goal will be to start with codes designed for a WTC-I$(t,u, q)$ and apply them to WTC-II$(t,u,q)$, without much loss in the measure of the operational parameters. 
Because of this, it is important to note that given a fixed channel state sequence, a WTC-I$(t,u, q)$ and WTC-II$(t,u,q)$ are equivalent.

Thus, in order to apply the WTC-I codes we have to negate the advantage introduced by the eavesdropper's channel state choice. Namely, we have to ensure that the wiretap channel type I code can hold for every possible channel state.
Our code transformation will accomplish this by permuting the order in which the symbols are transmitted, and sending information about the chosen permutation with a unique ``header'' code\footnote{In practice this code will only need to transmit $O(\log n)$ bits of information. We will implicitly assume the existence of such codes, and take for granted that the number of symbols needed to transmit this information reliably is $o(n)$.}.
The eavesdropper, whose choice of state sequence is fixed and independent of the encoder's output, will therefore not know a priori which state is being applied to which symbol. 

\begin{convention}
It will be necessary to reference sub-sequences of the output of a permutation.
To reference the $n(j-1)+1$ through $nj$-th symbols of $\upsilon^{-1}_{w}(y^{n\ell})$ would have us writing $\upsilon_{w}^{-1}(y^{n\ell})_{n(j-1)+1}^{nj}$. 
We feel that this unacceptable notation since we will only need to consider a very limited number of such sub-sequences. 
As an alternative we write
$$\upsilon_{w,j}^{-1} (y^{n\ell}) = \upsilon_{w}^{-1}(y^{n\ell})_{n(j-1)+1}^{nj} .$$
\end{convention} 

\begin{define}
Given codes $$\begin{array}{l r l r l} (F,\theta),  &F:\mcf{M} &\rightarrow \mcf{X}^n,&\theta : \mcf{Y}^n &\rightarrow \mcf{M} \\ (G,\varphi),&G: \mcf{ M}^* &\rightarrow \mcf{X}^{\kappa},&\varphi: \mcf{Y}^\kappa &\rightarrow \mcf{ M}^*\end{array}$$ and
a set of $n\ell$-symbol sequence permutations, $\mcf{V} = \{\upsilon_{i}\}_{i=1}^{|\mcf{M}|^*}$, the $\ell$-concatenate and permute code of $(F,\theta)$ with header $(G,\varphi)$ and permutations $\mcf{V}$ is defined as  
\begin{align}
&F^{(\ell)}_{G,\mcf{V}}(m^{\ell}) = \left[\upsilon_{W} \left( F(m_1),F(m_2),\dots,F(m_\ell)  \right), G(W) \right] \notag \\
&\theta_{\varphi,\mcf{V}}^{(\ell)}\left(y^{n\ell + \kappa}\right) \notag  \\
&\hspace{5pt}= \left[ \theta\left(\upsilon^{-1}_{\varphi\left( y_{n\ell+1}^{n\ell+\kappa}\right), 1 }(y^{n\ell}) \right) , 
%\theta\left(\upsilon^{-1}_{\varphi\left( y_{n\ell+1}^{n\ell+\kappa}\right), 2 }(y^{n\ell})  \right),
\dots, \theta\left( \upsilon^{-1}_{\varphi\left( y_{n\ell+1}^{n\ell+\kappa}\right), \ell }(y^{n\ell}) \right) \right] \notag
\end{align}
where $W$ is a uniform random variable over $\mcf{ M}^*$.
\end{define}

The transformed encoder can be viewed first as concatenating the output of $\ell$ uses of the original encoder $F$, to form a $n\ell$-symbol sequence. 
Next a permutation, chosen at random from the set $\mcf{V} = \{\upsilon_{i}\}_{i=1}^{|\mcf{ M}^*|}$, is applied to the $n\ell$-symbol sequence.
Finally the encoder uses $G$ to encode the value of the chosen permutation, and appends the encoded sequence to the $n\ell$-symbol sequence.
At the other end, the transformed decoder first uses $\varphi$ to decode the header which contains the information of which permutation was chosen. 
The decoder then applies the inverse permutation to the first $n\ell$ symbols, which in the absence of decoding error for the header, results in every consecutive non-overlapping sequence of $n$ symbols corresponding to a different use of the original code.   
Finally, the original decoder is applied to each consecutive block of $n$-symbols, in turn decoding each $m_j$, $j \in \{1,\dots,\ell\}$.

Before moving into the technical lemmas which make the main theorem possible, we wish to briefly discuss why the above aspects were necessary.
Of major importance is the permutation of the encoder outputs which suppresses the eavesdropper's ability to choose a specific state sequence to attack the code. 
To see this, consider the case where there exists only a small set of state sequences which are truly detrimental to the original code. 
Permuting the original code will also permute the set of detrimental state sequences, and with only a small probability will two randomly chosen permutations share a detrimental state sequence. 
Thus by choosing from a large number of possible permutations, it is unlikely that any state sequence chosen by the eavesdropper will be detrimental for the independently chosen permutation.
Even if the eavesdropper deciphers the header, thus learning which permutation was chosen, by then it is too late as it is likely there chosen state seqnece did not leak information. 

Thus the need for the permutation, but why the need to concatenate the codes?
Consider this: while a $(n,r,\epsilon,\delta)$ code for the WTC-I$(t,u, q)$ does provide
$$\mathbb{I}(Z^n,S^n;M) = \mathbb{I}(Z^n;M|S^n) \leq \delta,$$
it does not necessarily provide
$$ \mathbb{I}(Z^n;M|S^n,S^n\in \mcf{T}_{q}^n ) \leq \delta.$$
In fact, naively, it may be possible to have a code where $\mathbb{I}(Z^n;M|S^n ) = o(1)$ and $\mathbb{I}(Z^n;M|S^n,S^n\in \mcf{T}_{q}^n ) = n|\mcf{Z}|$, as long as $\mathbb{I}(Z^n;M|S^n,S^n\in \mcf{T}_{\tilde q}^n ) = 0$, for all $\tilde q  \in \mcf{P}_n(\mcf{S}) - \{q\}$.
This is problematic because in the WTC-II$(t,u,q)$ the eavesdropper has a fixed empirical distribution for the channel state sequence.
By concatenating multiple uses of the code together, and then applying permutations, the distribution of state symbols applied to each use of the code will appear closer to i.i.d. instead of chosen from a type set.

We now establish a series of technical lemmas relating the secrecy of the transformed code to the original, and the probability of a given state sequence being selected for the constituent codes.

%Before getting into the main theorem it will be important to prove a simple technical lemma relating the secrecy measure for the transformed code over the WTC-II to the secrecy measure of the original code over the WTC-I.
%The upcoming analysis will see us need to compare the secrecy measure of the original code to that of the transformed code. 
%This will be slightly problematic in terms of notation, as the secrecy measure is traditionally defined on random variables and thus will require us to define a set of random variables relating to the transformed code, and a set to the original code. 
%To that end we will us the following convention.
\begin{convention}
$\hat M, \hat Z^n,$ and $\hat S^n$ will be used to denote the message, eavesdropper's observation and the state sequence, respectively, for the original code, $(F,\theta)$, sent over a WTC-I$(t,u, q)$. 
While $M^\ell$, $Z^{n\ell+\kappa}$, and $S^{n\ell+\kappa}$ will be used to denote the message, eavesdropper's observation and eavesdropper's chosen state sequence when sent using the transformed code, $(F_{G,\mcf{V}}^{(\ell)} , \theta_{\varphi,\mcf{V}}^{(\ell)})$, over a WTC-II$(t,u,q)$.
\end{convention}

\begin{lemma}\label{lem:pita}
Let $(F,\theta)$ and $(G,\varphi)$ be a $(n,r,\epsilon_f,\delta)$ and $( \kappa , \psi,\epsilon_g,\infty)$ code, respectively, for a WTC-I$(t, u, q)$, and let $\mcf{V} = \{\upsilon_{i}\}_{i=1}^{2^{\kappa\psi}}$ be a collection of $n\ell$-symbol sequence permutations. 
Then for WTC-II$(t,u,q)$ and $(F^{(\ell)}_{G,\mcf{V}}, \theta^{(\ell)}_{\varphi,\mcf{V}})$-code,
\begin{align}
&\mathbb{I}(Z^{n\ell + \kappa};M^{\ell}|S^{n\ell+\kappa}  = s^{n\ell+\kappa}) \notag \\
&\hspace{40pt} \leq \sum_{j=1}^\ell \sum_{w = 1}^{2^{\kappa \psi}} \mathbb{I}(\hat Z^n;\hat M| \hat S^n = \upsilon_{w,j}^{-1}(s^{n\ell})) 2^{-\kappa \psi}.
\end{align}
\end{lemma}

\iftoggle{arxiv}
{\noindent For proof see Appendix~\ref{app:lem:pita}.}
{\begin{proofsketch}
This is essentially bookkeeping. 
Each use of code is independent, and the two channels are identical for a fixed state sequence. 
When the code has the inverse permutation applied to it, the channel state sequence being applied to the $j$-th code will be the $n(j-1)+1$ through $nj$-th symbols of the inverse permutation of the channel state sequence, that is $\upsilon^{-1}_{w,j}(s^{n\ell}).$
\end{proofsketch}}

We will also need to be able to determine the probability of a particular state sequence subject to a randomly chosen permutation.
\begin{lemma}\label{lem:oncurd}
Let $\Upsilon$ be uniform over the set of all $n$-symbol sequence permutations, and let $s^n \in \mcf{T}_{q}^n$, $q \in \mcf{P}_{n}(\mcf{S})$.
$$\Pr \left(\Upsilon(s^n) = \hat s^n \right) = \begin{cases}
\frac{q^n(\hat s^n)}{q^n(\mcf{T}_{q}^n)} &\text{ if } \hat s^n \in \mcf{T}_q^n \\
0 &\text{ o.w.} 
\end{cases}.$$
\end{lemma}
\iftoggle{arxiv}
{\noindent For proof see Appendix~\ref{app:lem:oncurd}.}
{\begin{proofsketch}
Permutations preserve the empirical distribution. 
Furthermore, each pair of elements in a type set has the same number of unique permutations that will transform one into the other. 
\end{proofsketch}}

%And for a final technical lemma we will need to be able to upper bound the probability of a state sequence having a particular sub-sequence. 
%This lemma is in part necessary because while a good code WTC-I$(t,u\cdot q)$ provides $$\mathbb{I}(\hat Z^n;\hat M|\hat S^n) < \delta,$$ it does not necessarily provide 
%$$\mathbb{I}(\hat Z^n;\hat M|\hat S^n \in \mcf{T}_{q}^{n}) < \delta.$$

%We will also need the following technical lemma which allows us to consider the probability of sequences drawn uniformly from a type set have a particular sub-sequence.

\begin{lemma}\label{lem:type_to_iid}
Let $n, \ell \in \mathbb{Z}^+$, and let $q \in \mcf{P}_{n\ell}(\mcf{S})$, where $\min_{s \in \mcf{S}} q(s) \geq 4 \sqrt{\frac{\ln (n(\ell-1))}{n(\ell-1)}}$.
\[
\frac{q^{n \ell } \left( \set{ s^{n \ell } : s^{n} = \hat s^{n}   } \cap \mcf{T}_{q}^{n \ell} \right)}{q^{n\ell}(\mcf{T}^{n\ell}_q)} \leq \mu q^{n}(\hat s^{n})
\]
where
\begin{align*}
\mu &= \sqrt{2} e^{-\frac{1}{4}} \frac{13}{11} \left( \frac{2 \pi \ell}{\ell-1} \right)^{\frac{|\mcf{S}|}{2}}.
\end{align*}
\end{lemma}

\iftoggle{arxiv}{\noindent For proof see Appendix~\ref{app:thm:cat:type_to_iid}.}{\begin{proofsketch}
First we observe that there exists a $\hat q \in \mcf{P}_{n(\ell-1)}$ such that $q^{n \ell } \left( \set{ s^{n \ell } : s^{n} = \hat s^{n}   } \cap \mcf{T}_{q}^{n \ell} \right) = q^{n}(\hat s^n) q^{n(\ell-1)}(\mcf{T}_{\hat q}^{n(\ell-1)})$.
With this in hand we use Stirling's approximation to obtain tight bounds on the probability of a type for a product measure. 
Finally we optimize those bounds to obtain the stated result.
\end{proofsketch}}

\begin{theorem}\label{thm:main}
Fix any $\lambda > 0$, $\ell \in \mathbb{Z}^+$ and let $(F,\theta)$ and $(G,\varphi)$ be a $(n,r,\epsilon_f,\delta)$ and $( \kappa , \psi,\epsilon_g,\infty)$ code, respectively, for a WTC-I$(t, u, q)$ where $q \in \mcf{P}_{n\ell + \kappa}(\mcf{S})$, $\min_{s \in \mcf{S}} q(s) \geq 4 \sqrt{\frac{\ln (n(\ell-1))}{n(\ell-1)}} + \frac{\kappa}{n\ell}$. 
Given independent random variables $\{\Upsilon_{i}\}_{i=1}^{2^{\kappa\psi}}$ uniformly distributed over the set of $n\ell$-symbol sequence permutations,
$$\Pr \left( \Upsilon^{2^{\kappa\psi}} \notin \mcf{V}^* \right)<  \ell |\mcf{S}|^{n\ell}  \exp{\left(- 2^{\kappa\psi+1} \left(\frac{\lambda}{ n |\mcf{Z}| }\right)^2 \right)},$$
where $\mcf{V}^*$ is the collection of $\mcf{V} = \{\upsilon_i\}_{i=1}^{2^{\kappa\psi}}$ such that $(F^{(\ell)}_{G,\mcf{V}}, \theta^{(\ell)}_{\varphi,\mcf{V}})$ is a  
$$\left(n\ell+\kappa, r \left[1 - \frac{\kappa}{n\ell + \kappa}\right] , \ell\epsilon_f + \epsilon_g, \ell  \left[ \mu e^{\kappa/\ell} \delta + \lambda \right] \right)$$
WTC-II$(t,u,q)$ code, where $\mu$ is from Lemma~\ref{lem:type_to_iid}.

%If each $\{\upsilon_{i}\}_{i=1}^{2^{\kappa\psi}}$ is independently chosen uniformly from the set of all $n\ell$-symbol sequence permutations, then with probability 
%$$> 1- 2 |\mcf{S}|^{n\ell} \ell \exp{\left(- 2^{\kappa\psi+1} \left(\frac{\lambda \mu 2^{\frac{\kappa}{\ell}} \delta}{ n |\mcf{Z}| }\right)^2 \right)},$$
%a set of permutations, $\mcf{V},$ will be chosen such that $(F^{(\ell)}_{G,\mcf{V}}, \theta^{(\ell)}_{\varphi,\mcf{V}})$ is a
%$$\left(n\ell+\kappa, r \left(1 - \frac{\kappa}{n\ell + \kappa}\right) , \ell\epsilon_f + \epsilon_g, (1+\lambda) \ell \mu 2^{\frac{\kappa}{\ell}} \delta\right)$$
%WTC-II$(t,u,q)$ code.

\end{theorem}

\begin{remark}
Notice that $\psi$ is a coding rate and should be close to the capacity of the channel for large values of $\kappa$. 
Thus setting $\lambda = n^{-2} $, choosing $ n^8 |\mcf{Z}|^2/2$ permutations and concatenating 
$$\ell =  \frac{-1 + \logt (n^4 |\mcf{Z}|^2) - \logt 2 \lambda^2}{\psi}  $$ codes yields a
$$\approx \left(n\ell, r, \epsilon_f \ell, (\delta + n^{-2} ) \ell    \right)\text{WTC-II code}$$
with probability $\approx 1 - e^{-n^2}.$ 
Therefore, for\footnote{This is the minimum value of $\delta$ needed to assure a semantic security sub code, see~\cite{wiese2018semantic}.} $\delta \geq n^{-2}$ and small $\epsilon_f$, the new code produced for the WTC-II sends nearly the same amount of information, in nearly the same number of symbols, with nearly the same error probability, and nearly the same information leakage as $\ell$ uses of the original code would over the WTC-I.

\end{remark}

\begin{remark}
For future work we hope to derive an approximation lemma, similar to that used in~\cite{ahlswede1978elimination}, thus eliminating the need for the bound on $q(s).$ 

\end{remark}

\begin{IEEEproof}
Clearly, regardless of chosen $\mcf{V}$, $(F^{(\ell)}_{G,\mcf{V}}, \theta^{(\ell)}_{\varphi,\mcf{V}})$ will use $n\ell +\kappa$ symbols, to transmit a message from a set of size $|\mcf{M}|^{\ell} = 2^{\ell n r}$ yielding rate
\begin{align}
\frac{n  \ell r}{n \ell + \kappa}  = \left( 1 - \frac{\kappa}{n\ell + \kappa} \right) r .   
\end{align}
Furthermore regardless of permutations chosen (since each is invertible) the probability of error must be less than $\epsilon_g + \ell \epsilon_f$ by the union bound, accounting for the single use of $(G,\varphi)$ (probability of error $\epsilon_g$) and the $\ell$ uses of $(F,\theta)$ (probability of error $\epsilon_f$). 

What remains is to calculate the secrecy measure. 
%Let $W$ be the random variable denoting which of the $2^{\kappa \psi}$ permutations is chosen and transmitted.
%Note the support of $\mcf{W} = \{1,\dots,2^{\kappa\psi}\}.$
For now fix $S^{n\ell + \kappa} = s^{n\ell+\kappa},$ and observe that if  $\Upsilon^{2^{\kappa\psi}} = \{ \upsilon\}_{i=1}^{2^{\kappa\psi}}$ then
\begin{align}
&\mathbb{I}(Z^{n\ell + \kappa};M^{\ell}|S^{n\ell+\kappa}  = s^{n\ell+\kappa}) \notag \\
&\hspace{40pt} \leq \sum_{j=1}^\ell \sum_{w = 1}^{2^{\kappa \psi}} \mathbb{I}(\hat Z^n;\hat M| \hat S^n = \upsilon_{w,j}^{-1}(s^{n\ell})) 2^{-\kappa \psi}
\end{align}
by Lemma~\ref{lem:pita}.

Set $V_{w,j} =  \mathbb{I}_{\Upsilon_{w}}\left(\hat Z^{n}; \hat M \middle| \hat S^n = \Upsilon_{w,j}^{-1}(s^{n\ell})\right),$ and note for later that given a fixed $j$ the random variables $\{V_{w,j}\}_{w =1}^{2^{\kappa\psi}}$ are independent since $\{\Upsilon_w\}_{w =1}^{2^{\kappa\psi}}$ are independent and $V_{w,j}$ is a deterministic function of $\Upsilon_w.$
With this notation, the probability of choosing a set of permutations such that $\mathbb{I}(Z^{n\ell + \kappa};M^{\ell}|S^{n\ell+\kappa}  = s^{n\ell+\kappa}) >  \ell  e^{{\kappa}/{\ell}} \mu \delta  + \ell \lambda$ is bounded by
\begin{align}
&\Pr \left( \sum_{j=1}^{\ell} \sum_{w=1}^{2^{\kappa\psi}} V_{w,j} 2^{-\kappa \psi} \geq  \ell e^{\kappa/\ell} \mu \delta + \ell \lambda  \right) \notag\\
&\hspace{20pt} \leq \sum_{j=1}^{\ell}\Pr \left(  \sum_{w=1}^{2^{\kappa\psi}} V_{w,j} 2^{-\kappa \psi} \geq e^{\kappa/\ell} \mu \delta + \lambda  \right) .
\end{align}
Hence, if \begin{align}
\mathbb{E}\left[ \sum_{w=1}^{2^{\kappa\psi}} 2^{-\kappa\psi} V_{w,j} \right] \leq e^{\kappa/\ell} \mu \delta, \label{eq:main:exp-1}
\end{align}
which we shall return to later, then 
\begin{align}
&\Pr \left( \sum_{j=1}^{\ell} \sum_{w=1}^{2^{\kappa\psi}} V_{w,j} 2^{-\kappa \psi} \geq  \ell e^{\kappa/\ell} \mu \delta + \ell \lambda  \right) \notag\\
%&\leq \sum_{j=1}^{\ell}\Pr \left(  \sum_{w=1}^{2^{\kappa\psi}} V_{w,j} 2^{-\kappa \psi} \geq 2^{\kappa/\ell} \mu \delta + \lambda  \right) \\
&\leq \sum_{j=1}^{\ell}\Pr \left(  \sum_{w=1}^{2^{\kappa\psi}} V_{w,j} 2^{-\kappa \psi}  - \mathbb{E}\left[ \sum_{w=1}^{2^{\kappa\psi}} 2^{-\kappa\psi} V_{w,j} \right]\geq  \lambda  \right) \\
%&\leq \sum_{j=1}^{\ell} \exp{\left(-2^{\kappa\psi+1} \left( \frac{\lambda}{n |\mcf{Z}|} \right)^2\right)} \\
&\leq \ell \exp{\left(-2^{\kappa\psi+1} \left( \frac{\lambda}{n |\mcf{Z}|} \right)^2\right)} \label{eq:main:pr:1}
\end{align}
where~\eqref{eq:main:pr:1} follows from Hoeffding's inequality\footnote{Note that $0\leq V_{w,j} \leq n |\mcf{Z}|$ by the non-negativity of mutual information and that $ \mathbb{I}_{\Upsilon_{w}}\left(\hat Z^{n}; \hat M \middle| \hat S^n = \Upsilon_{w,j}^{-1}(s^{n\ell})\right) \leq \mathbb{H}(\hat Z^n |\hat S^n = \Upsilon_{w,j}^{-1}(s^{n\ell})) \leq n |\mcf{Z}|$.}.
Since the choice of $s^{n\ell+\kappa}$ was arbitrary, the probability of selecting a set of permutations such that $$\max_{s^{n\ell+\kappa} \in \mcf{T}_{q}^{n\ell+\eta}} \mathbb{I}(Z^{n\ell + \kappa};M^{\ell}|S^{n\ell+\kappa}  = s^{n\ell+\kappa}) > \ell e^{\kappa/\ell}\mu \delta + \ell \lambda$$ must be less than 
$$\ell |\mcf{S}|^{n\ell + \kappa} \exp{\left(-2^{\kappa\psi+1} \left( \frac{\lambda}{n |\mcf{Z}|} \right)^2\right)} $$
by Equation~\eqref{eq:main:pr:1} combined with the union bound to account for all possible state sequence choices.

To finish the proof we return to show Equation~\eqref{eq:main:exp-1}.
Begin by writing the LHS of Equation~\eqref{eq:main:exp-1} as
\begin{align}
\sum_{\hat s^n} \Pr \left( \Upsilon_{w,j}^{-1}(s^{n\ell}) = \hat s^n\right) \mathbb{I}(\hat Z^n;\hat M|\hat S^n = \hat s^n) \label{eq:main:exp1}
\end{align}
which can be done since $V_{w,j} = \mathbb{I}_{\Upsilon_w}(\hat Z^n;\hat M|\hat S^n= \Upsilon^{-1}_{w,j}(s^{n\ell})),$ for some arbitrary $s^{n\ell}$. 
Now it is clear that~\eqref{eq:main:exp1} equals 
\begin{align}
\sum_{\hat s^n} \frac{ \tilde q^{n\ell} ( \{s^{n\ell}: s^n = \hat s^n\} \cap \mcf{T}_{\tilde q}^{n\ell})}{\tilde q^{n\ell}(\mcf{T}_{q}^{n\ell})} \mathbb{I}(\hat Z^n;\hat M|\hat S^n = \hat s^n), \label{eq:main:exp2}
\end{align}
where $\tilde q$ is the empirical distribution of $s^{n\ell},$ by Lemma~\ref{lem:oncurd}.
But the difference between the empirical distributions of $s^{n\ell+\kappa}$ and $s^{n\ell}$ must be small. 
In fact each $s \in \mcf{S}$ occurs $q(s) (n\ell + \kappa)$ times in $s^{n\ell+\kappa}$, and hence can occur in $s^{n\ell}$ at most $q(s) (n\ell + \kappa)$ times and at least $q(s)(n\ell+\kappa) - \kappa$ times. 
Therefore we have that 
\begin{align}
\tilde q(s) \leq \frac{q(s) (n\ell + \kappa)}{n\ell } \text{ or } \frac{\tilde q(s)}{q(s)} \leq 1 + \frac{\kappa}{n\ell}, \label{eq:main:exp2b}
\end{align}
which will be of use shortly, and 
\begin{align}
\tilde q(s) \geq \frac{q(s) (n\ell + \kappa) - \kappa }{n\ell } >  q(s) - \frac{\kappa}{n\ell} \geq 4\sqrt{\frac{\ln(n(\ell-1))}{n(\ell-1)}}\label{eq:main:exp2c}
\end{align}
by the assumptions on $q$ in the theorem statement. 
We can upper bound Equation~\eqref{eq:main:exp2} with
\begin{align}
\sum_{\hat s^n} \mu \tilde q^{n}(\hat s^n)  \mathbb{I}(\hat Z^n;\hat M|\hat S^n = \hat s^n), \label{eq:main:exp3}
\end{align}
using Lemma~\ref{lem:type_to_iid}, in light of Equation~\eqref{eq:main:exp2c}.
Furthermore, using Equation~\eqref{eq:main:exp2b} it follows that
\begin{align}
\tilde q^n(\hat s^n) \leq q^n(\hat s^n) \left( 1 + \frac{\kappa}{n\ell} \right)^{n} %=q^n(\hat s^n) e^{n \ln \left( 1 + \frac{\kappa}{n\ell} \right)} 
\leq q^n(\hat s^n) e^{\kappa/\ell} .\label{eq:main:exp3b}
\end{align}
Combining Equations~\eqref{eq:main:exp3} and~\eqref{eq:main:exp3b} yields 
\begin{align}
\mu e^{\kappa/\ell} \sum_{\hat s^n}   q^{n}(\hat s^n)  \mathbb{I}(\hat Z^n;\hat M|\hat S^n = \hat s^n) \label{eq:main:exp4}
\end{align}
as an upper bound of $\mathbb{E}\left[ \sum_{w=1}^{2^{\kappa\psi}} 2^{-\kappa\psi} V_{w,j} \right].$
Finally, observe that $\hat S^n$ is distributed $q^n$ in the WTC-I$(t,u, q)$, and thus 
\begin{align}
\mu e^{\kappa/\ell} \sum_{\hat s^n}   q^{n}(\hat s^n)  \mathbb{I}(\hat Z^n;\hat M|\hat S^n = \hat s^n)
&= \mu e^{\kappa/\ell} \mathbb{I}(\hat Z^n; \hat M|\hat S^n) \notag \\
%&\leq \mu e^{\kappa/\ell} \mathbb{I}(\hat Z^n,\hat S^n; \hat M) \\
&\leq  \mu e^{\kappa/\ell} \delta ,\label{eq:main:exp4b}
\end{align}
proving Equation~\eqref{eq:main:exp-1}.

\end{IEEEproof}

\bibliographystyle{IEEEtran}
\bibliography{this} 

\iftoggle{arxiv}{\appendices

\section{Proof of Lemma~\ref{lem:pita}}\label{app:lem:pita}

\begin{IEEEproof} 
This Lemma is derived as follows
\begin{align}
&\mathbb{I}(Z^{n\ell+\kappa};M^{\ell}|S^{n\ell + \kappa}) \notag\\
& \leq \mathbb{I}(Z^{n\ell+\kappa},W;M^{\ell}|S^{n\ell + \kappa}=s^{n\ell+\kappa}) \label{eq:pita1}\\
&= \mathbb{I}(Z^{n\ell};M^{\ell}|S^{n\ell}=s^{n\ell+\kappa},W) \label{eq:pita2}\\
&= \sum_{w=1}^{2^{\kappa \psi}} 2^{-\kappa \psi} \mathbb{I}(Z^{n\ell};M^{\ell}|S^{n\ell }=s^{n\ell},W=w) \label{eq:pita3}\\
&= \sum_{w=1}^{2^{\kappa \psi}} \sum_{j=1}^\ell 2^{-\kappa \psi} \mathbb{I}(\hat Z^{n};\hat M |\hat S^{n}=v_{w,j}^{-1}(s^{n\ell})) \label{eq:pita4}
\end{align}
where~\eqref{eq:pita2} is because\footnote{Recall that $W, M^\ell$ and $S^n$ are chosen independently.} $\mathbb{I}(W;M^{\ell}|S^{n\ell+\kappa} = s^{n\ell+\kappa}) = 0$ while the last $\kappa$ symbols are independent given header information $W$, and~\eqref{eq:pita4} is because each use of the code is independent and the channel uses are independent given the state sequence which for the $j$-th code corresponds to the $n(j-1)+1$ through $nj$-th symbols of the inverse permutation of the channel state sequence.

\end{IEEEproof}

\section{Proof of Lemma~\ref{lem:oncurd}}\label{app:lem:oncurd}
\begin{IEEEproof}
Let $a^n,b^n,c^n$ be (not necessarily distinct) sequences of type $q \in \mcf{P}_n(\mcf{S}).$
Suppose in hopes of a contradiction that there existed more permutations in $\mcf{U}^*(b^n) \defn \{ \upsilon: \upsilon(a^n) = b^n\}$ than in $\mcf{U}^*(c^n).$
Now, clearly, there exists at least one permutation $\tilde \upsilon$ such that $\tilde\upsilon(b^n) = c^n.$
Furthermore $\tilde \upsilon( \upsilon(a^n)) = c^n$  for all $\upsilon \in \mcf{U}^*(b^n).$
But this implies $\tilde \upsilon(\mcf{U}^*(b^n)) \subseteq \mcf{U}^*(c^n)$ and thus $|\mcf{U}^{*}(b^n)|\leq |\mcf{U}^*(c^n)|$ since all permutations are invertible functions.
This is a contradiction, and therefore $|\mcf{U}^{*}(b^n)|$ is equal for all $b^n \in \mcf{T}^n_q.$

The above contradiction implies that if $\Upsilon$ is uniformly chosen from the set of all permutations then
$$\Pr \left(\Upsilon(a^n) = b^n \right) = \frac{1}{|\mcf{T}_q^n|} = \frac{q^n(b^n)}{q^n(\mcf{T}_q^n)}.$$

\end{IEEEproof}

\section{Proof of Lemma~\ref{lem:type_to_iid}}\label{app:thm:cat:type_to_iid}

\begin{IEEEproof}

%\phi = n -n_1 -> n\ell

Begin by observing that if the empirical distribution of $\hat s^{n}$ is $\tilde q$, then 
\begin{equation}\label{eq:thm:cat:strans}
\set{s^{n\ell} : s^{n} = \hat s^{n}} \cap \mcf{T}_q^{n\ell} = \set{s^{n\ell} : s^{n} = \hat s^{n} , s_{n+1}^{n\ell } \in \mcf{T}^{n\ell}_{\hat q} },
\end{equation}
where $\hat q \in \mcf{P}(\mcf{S})$ is the distribution such that 
\begin{equation}\label{eq:thm:cat:qtrans}
\hat q \frac{n\ell -n }{n\ell } + \tilde q \frac{n}{n\ell } = \hat q \frac{\ell -1 }{\ell } + \tilde q \frac{1}{\ell }  = q.
\end{equation}
From Equation~\eqref{eq:thm:cat:strans} that
\begin{equation}\label{eq:thm:cat:strans_cons}
\frac{q^{n\ell}\left( \set{s^{n\ell} : s^{n} = \hat s^{n}} \cap \mcf{T}_q^{n\ell} \right)}{q^{n\ell}(\mcf{T}^{n\ell}_q)} = \frac{q^{n}(\hat s^{n}) q^{n(\ell-1)} ( \mcf{T}_{\hat q}^{n(\ell-1)}) }{q^{n\ell}(\mcf{T}^{n\ell}_q)}
\end{equation}
directly follows due to $q^n$ being a product measure. Notice now that the RHS of Equation~\eqref{eq:thm:cat:strans_cons} takes the form of $q^{n}(\hat s^{n})$ multiplied $q^{n(\ell-1)}(\mcf{T}^{n(\ell-1)}_{\hat q}) / q^{n\ell}(\mcf{T}_q^{n\ell})$, and this multiplier can be computed by calculating the probabilities of specific types. As a first attempt it is tempting to use the traditional bounds (such as~\cite[Lemma~2.6]{CK}, $(k+1)^{-\left|\mcf{S}\right|} 2^{-k \mathbb{D}(\hat q|| q)} \leq q^k ( \mcf{T}^k_{\hat q} ) \leq 2^{- k \mathbb{D}(\hat q || q)}$), but they are far too loose and result in a multiplier which is polynomial with $k$.
Instead applying the stricter bounds derived from Stirling's approximation, namely
\begin{align}
 q^k ( \mcf{T}^k_{\hat q} ) &= \left( 1 + \zeta \right) \frac{1}{\sqrt{2 \pi k}^{\left|\mcf{S}\right|-1}\sqrt{\prod_{i=1}^{\left|\mcf{S}\right|} \hat q(i) } } 2^{-k \mathbb{D}(\hat q || q) }  \label{eq:thm:cat:sa_type},
\end{align}
where $|\zeta| \leq \frac{1}{12k}$
results in
\begin{align}
&\frac{ q^{n(\ell-1)} ( \mcf{T}_{\hat q}^{n(\ell-1)}) }{q^{n\ell}(\mcf{T}^{n\ell}_q)} 
\notag \\
&\leq \hat \zeta   \prod_{i\in \mcf{\hat S}} \left( \sqrt{  \frac{q(i)}{\hat q(i)}}\right) \prod_{i\in \mcf{S} - \mcf{\hat S}} \left( \sqrt{  n (\ell-1)  q(i)} \right) 2^{-n(\ell-1) \mathbb{D}(\hat q||q)}, \label{eq:thm:cat:mult_app}
\end{align}
where $\mcf{\hat S} \subset \mcf{S}$ are the indices $i$ for which $\hat q(i) \neq 0$, and 
\begin{align*}
\hat \zeta = \frac{1 + \frac{1}{12n(\ell-1)  }}{1 - \frac{1}{12n\ell }} \sqrt{2 \pi }^{\left|\mcf{S}\right|-\left|\mcf{\hat S}\right| }  \sqrt{\frac{\ell}{\ell-1}}^{\left|\mcf{ S}\right| -1 }&\hspace{-10pt}\leq \frac{13}{11} \left( \frac{2 \pi \ell}{\ell-1} \right)^{\frac{|\mcf{S}|}{2}}.
\end{align*} 
Equation~\eqref{eq:thm:cat:mult_app} in turn has upper bound
\begin{align}
&\hat \zeta \prod_{i\in \mcf{\hat S}} \left( \sqrt{ \frac{q(i)}{q(i) - \Delta(i)} }\right)  \prod_{i\in \mcf{S} - \mcf{\hat S}} \left( \sqrt{  q(i) n(\ell-1) } \right) \notag \\
&\hspace{30pt} \cdot e^{-\frac{n(\ell-1)}{2 } \left(\sum_{i\in \mcf{S}} \left|\Delta(i) \right| \right)^2 } \notag \\
&\leq \hat \zeta \prod_{i\in \mcf{\hat S}} \left( \sqrt{ \frac{q(i)}{q(i) - \Delta(i)} }\right)  \prod_{i\in \mcf{S} - \mcf{\hat S}} \left( \sqrt{  q(i)n(\ell-1)} \right) \notag \\
&\hspace{30pt} \cdot e^{-\frac{n(\ell-1)}{2 } \sum_{i\in \mcf{S}} \left|\Delta(i) \right|^2 } \label{eq:thm:cat:pins} ,
\end{align}
where
%\footnote{From this we obtain $\hat q(i) = q(i) - \Delta(i)/(\ell-1)$ by Equation~\eqref{eq:thm:cat:strans_cons}.} 
$\Delta(i) =  q(i) - \hat q(i) $, due to an application of Pinsker's inequality. The advantage of Equation~\eqref{eq:thm:cat:pins} is that it can be written
\begin{equation}\label{eq:thm:cat:pins_alt}
\hat \zeta  \prod_{i\in \mcf{S}} \tau_i ( \Delta(i) ),
\end{equation}
where 
\begin{align}
\tau_{i}(x) &\defn \begin{cases} 
e^{-\frac{n(\ell-1)}{2 } x^2 } \sqrt{ \frac{ q(i)}{ q(i) - x} }  &\text{ if } i \in \mcf{\hat S}   \\ 
& \hspace{-80pt} \text{ and } x\in \left( - \infty,  q(i) - \frac{1}{n(\ell-1)}  \right]\\
e^{-\frac{n(\ell-1)}{2} q^2(i)} \sqrt{q(i)n (\ell-1)} &\text{ if } i \in \mcf{S} - \mcf{\hat S} \\
0 &\text{ o.w.}
\end{cases}\label{eq:thm:cat:f_equation},
\end{align}
and $\tau_i(x)$ has at most two local maximums for each $i$. 
Indeed, if $i \in \mcf{S}- \mcf{\hat S}$ then the maximum is \begin{align}
e^{-\frac{n(\ell-1)}{2} q^2(i)}\sqrt{q(i)n(\ell-1)}. \label{eq:thm:cat:i0fmax}
\end{align} 
On the other hand if $i \in \mcf{\hat S}$, then one of the maximums of $\tau_i(x)$ occurs at 
\begin{equation} \label{eq:thm:cat:max_x0}
x = q(i) - \frac{1}{n(\ell-1)},
\end{equation} 
and the other at\footnote{Only valid if $q^2(i) n(\ell-1) > 2$.}
%\begin{equation}\label{eq:thm:cat:max_x}
%x = \frac{\phi q(i)}{2 n_1} - \frac{1}{2} \sqrt{ \left(\frac{\phi q(i)}{n_1}\right)^2 - \frac{2}{\phi}},
%\end{equation}
\begin{equation}\label{eq:thm:cat:max_x}
x =   \left( \frac{1}{2}  -  \frac{1}{2}\sqrt{ 1 - \frac{2}{q^2(i) n (\ell-1)}}\right) q(i),
\end{equation}
which can be obtained, of course, by setting the derivative of Equation~\eqref{eq:thm:cat:f_equation} to zero, and solving\footnote{There is also a minimum at $\left( \frac{1}{2}  +  \frac{1}{2}\sqrt{ 1 - \frac{2}{q(i)^2 n (\ell-1)}}\right) q(i)$. } for $x$, and then checking for asymptotes and boundary points. 
Evaluating $\tau_i$, for $i \in \mcf{\hat S}$, at these local maximums gives values of 
\begin{equation}\label{eq:thm:cat:i1fmax0}
e^{-\frac{n(\ell-1)}{2 } \left( q(i) - \frac{1}{n(\ell-1)} \right)^2 } \sqrt{  q(i) n (\ell-1) }
\end{equation}
and
%\begin{equation}\label{eq:thm:cat:fmax}
%e^{-\frac{q^2(i)n(\ell-1)}{8 } \left(  1  -  \sqrt{ 1 - \frac{2}{q^2(i) n (\ell-1)}} \right)^2 } \sqrt{ \frac{2}{1 + \sqrt{ 1 - \frac{2}{q^2(i) n (\ell-1)}} }}
%\end{equation}
%\begin{align}
%&e^{-\frac{q^2(i)n(\ell-1)}{8 } \left(  1  -  \sqrt{ 1 - \frac{2}{q^2(i) n (\ell-1)}} \right)^2 } \notag\\
%&\hspace{20pt} \cdot \sqrt{q^2(i) n (\ell-1) \left( 1 - \sqrt{ 1 - \frac{2}{q^2(i) n (\ell-1)}}\right) } 
%    \label{eq:thm:cat:fmax}
%\end{align}
\begin{align}
&e^{-\frac{n(\ell-1)}{8 } \left(  q(i)  -  \sqrt{ q^2(i) - \frac{2}{ n (\ell-1)}} \right)^2 } \notag\\
&\hspace{20pt} \cdot \sqrt{q(i) n (\ell-1) \left( q(i) - \sqrt{ q^2(i) - \frac{2}{ n (\ell-1)}}\right) } 
    \label{eq:thm:cat:i1fmax1}
\end{align}
respectively. 
For these maximums, it is easy to see that for equal values of $q(i)$ that the maximum in~\eqref{eq:thm:cat:i0fmax} is less than the maximum in~\eqref{eq:thm:cat:i1fmax0} since we have assumed 
$$q(s) > 4 \sqrt{\frac{\ln (n(\ell-1))}{n(\ell-1)} } $$
for all $s \in \mcf{S}$. 
Furthermore, from basic calculus we see that Equation~\eqref{eq:thm:cat:i1fmax0} is maximized when $q(i)$ is as small as possible.
Hence, we can obtain an upper bound on Equations~\eqref{eq:thm:cat:i0fmax} and~\eqref{eq:thm:cat:i1fmax0} as follows
\begin{align}
&e^{-\frac{n(\ell-1)}{2 } \left( q(i) - \frac{1}{n(\ell-1)} \right)^2 } \sqrt{  q(i) n (\ell-1) }\notag\\
&\leq e^{-\frac{n(\ell-1)}{2 } \left(4 \sqrt{\frac{\ln (n(\ell-1))}{n(\ell-1)} } - \frac{1}{n(\ell-1)} \right)^2 }  2 \left( n (\ell-1) \ln \left[n(\ell-1)\right]  \right)^{\frac{1}{4}} \\
&\leq e^{-\frac{n(\ell-1)}{2 } \left(2 \sqrt{\frac{\ln (n(\ell-1))}{n(\ell-1)} } \right)^2 }  2 \left( n (\ell-1) \ln \left[n(\ell-1)\right]  \right)^{\frac{1}{4}} \\
&= 2\left(\frac{\ln \left[ n (\ell-1) \right]}{\left( n \left( \ell -1 \right) \right)^7}\right)^{\frac{1}{4}} < 1, 
\end{align}
for all $n(\ell-1) \geq 2.$
On the other hand $q(s) = \sqrt{\frac{2}{n(\ell-1)}}$ maximizes Equation~\eqref{eq:thm:cat:i1fmax1}, which once again follows from basic calculus.
Evaluating~\eqref{eq:thm:cat:i1fmax1} at $q(s) = \sqrt{\frac{2}{n(\ell-1)}}$, amazingly, yields $\sqrt{2}e^{-\frac{1}{4}}.$

\end{IEEEproof}}{ }

\end{document}